\documentclass[aps,prl,twocolumn,floatfix,showpacs,preprintnumbers,amsmath,amssymb,nofootinbib,groupedaddress]{revtex4-1}
\usepackage{color}
\usepackage{graphicx}
\usepackage{dcolumn}    
\usepackage{multirow}
\usepackage{bm}         
\usepackage{sidecap}
\usepackage{hyperref}   
\usepackage{xcolor}
\usepackage{amsmath}
\usepackage{color,colortbl}
\usepackage{booktabs} 
\usepackage{hhline}
\usepackage{ulem}

\usepackage{hyperref} 
\definecolor{dark-red}{rgb}{0.,0.,0}
\definecolor{dark-blue}{rgb}{0.,0.,1}
\definecolor{medium-blue}{rgb}{0,0,1}
\definecolor{gray}{rgb}{0.85,0.85,0.85}

\hypersetup{
    colorlinks, linkcolor={dark-red},
    citecolor={dark-blue}, urlcolor={medium-blue}
}

\begin{document}

\title{Optimizing the relativistic energy density functional with nuclear ground state and collective excitation properties}

\author{E. Y{\"{u}}ksel}
\email{eyuksel@yildiz.edu.tr}
\affiliation {Department of Physics, Faculty of Science, University of Zagreb, Bijeni\v{c}ka c. 32, 10000 Zagreb, Croatia}
\affiliation {Department of Physics, Yildiz Technical University, 34220
Esenler, Istanbul, Turkey}

\author{T. Marketin}
\altaffiliation[Current affiliation: ]{Ericsson Nikola Tesla d.d., Krapinska 45, 10000 Zagreb, Croatia}
\affiliation {Department of Physics, Faculty of Science, University of Zagreb, Bijeni\v{c}ka c. 32, 10000 Zagreb, Croatia}

\author{N.Paar}
\email{npaar@phy.hr}
\affiliation {Department of Physics, Faculty of Science, University of Zagreb, Bijeni\v{c}ka c. 32, 10000 Zagreb, Croatia}

\date{\today} 

%
\begin{abstract}
We introduce a new relativistic energy density functional constrained by the ground state properties of atomic nuclei along with the isoscalar giant monopole resonance energy and dipole polarizability in $^{208}$Pb. A unified framework of the relativistic Hartree-Bogoliubov model and random phase approximation based on the relativistic density-dependent point coupling interaction
is established in order to determine the DD-PCX parameterization by $\chi^2$ minimization. This procedure is supplemented with the co-variance analysis in order to estimate statistical uncertainties in the model parameters and observables. The effective interaction DD-PCX accurately describes the nuclear ground state properties including the neutron-skin thickness, as well as the isoscalar giant monopole resonance excitation energies and dipole polarizabilities. The implementation of the experimental data on nuclear excitations allows constraining the symmetry energy close to the saturation density, and the incompressibility of nuclear matter by using genuine observables on finite nuclei in the $\chi^2$ minimization protocol, rather than using pseudo-observables on the nuclear matter, or by relying on the ground state properties only, as it has been customary in the previous studies.

\end{abstract}



\maketitle
Solving the quantum many-body problem of strongly interacting nucleons
represents one of the fundamental challenges not only for understanding the phenomena of nuclear
structure and dynamics, but also for various applications of astrophysical
relevance, e.g., modeling the stellar evolution, supernova explosion, the properties 
of compact stars, the synthesis of chemical elements in the universe, etc. Among a variety of
theoretical frameworks to address this problem, the nuclear energy density functional (EDF)
represents a unified approach to study quantitatively static and dynamic properties of finite nuclei along the nuclide map~\cite{nik11,paar07} as well as the equation of state of nuclear matter~\cite{roc18}. 
A considerable progress has been achieved in constructing
and optimizing the phenomenological EDFs, both in non-relativistic ~\cite{klup09,gog09,kor10} and
relativistic~\cite{nik11,lal05,nik08,zha10,pie14} frameworks. Recently, the construction of the EDFs
was also inspired by {\it ab initio} calculations~\cite{dobaczewski16} 
and effective field theories~\cite{bonnard18}. As pointed out in Ref. \cite{shen18},
strengths of the tensor forces guided by {\it ab initio} relativistic Brueckner-Hartree-Fock calculations
can also be used as a guide for the future {\it ab initio} derivations of the EDFs.
At present, only the phenomenological EDFs provide a level of accuracy required to quantitatively describe the nuclear properties across the whole nuclide map.

So far, the EDFs have mainly been parametrized with the experimental data
on the ground state properties of nuclei. These observables alone
are often not enough to constrain the
effective interaction completely, especially its isovector channel, thus
the protocols to determine the EDF's parameters often included constraints on the 
pseudo-observables on the nuclear matter properties. The neutron skin thickness $r_{np}$, isovector dipole
excitations in nuclei and neutron star mass and radii represent some of the possible observables that could be used to probe the isovector channel of the EDFs \cite{roc18,bal16,piek14}. However, the data on $r_{np}$ are often model
dependent, and the most recent data from parity violating electron
scattering experiment (PREX) on $^{208}$Pb~\cite{prex12} have large uncertainties, while
the neutron star mass-radius data are still rather limited.
An alternative observable to improve the performance of the EDF's isovector sector 
is the dipole polarizability ($\alpha_{D}$), which is proportional to the
inversely energy-weighted sum of the isovector dipole excitation in
a nucleus. Recently, it has been represented as an observable
strongly correlated with the symmetry energy parameters of the nuclear 
equation of state \cite{roc15,rein10}. The dipole polarizability was also measured in several nuclei, including $^{48}$Ca, $^{68}$Ni, $^{120}$Sn and $^{208}$Pb \cite{bir17,ros13,has15,tami11}.
So far, mainly due to
computational difficulties, the dipole polarizability was not
employed to constrain the EDFs directly; instead,
already established EDFs were tested for their performance in
reproducing the experimental data on $\alpha_D$ \cite{roc18}. 
On the other side, the isoscalar giant monopole resonance (ISGMR) energy in nuclei also represents an important observable to probe the isoscalar channel of the EDFs and incompressibility of nuclear matter $K_{0}$~\cite{gar18}. To date, only in Ref.~\cite{pie14} the ISGMR was used directly in the optimization of an EDF (relativistic FSUGold2 interaction) along with the properties of finite
nuclei and neutron stars.

The purpose of this work is to establish the first unified
framework based on the EDF, that allows to constrain an effective interaction not only
by the experimental data on nuclear ground state properties, but also by a direct implementation of 
the measured properties of collective nuclear phenomena: the ISGMR excitation energy 
and the dipole polarizability.  
In this way, the properties of nuclei and the nuclear equation of state; incompressibility 
of nuclear matter and the symmetry energy around the saturation density will be constrained directly
by the experimental data on finite nuclei.


In the framework of the relativistic nuclear energy density functional, the nuclear ground-state
density and energy are determined by the self-consistent solution of relativistic single-nucleon
Kohn-Sham equations~\cite{koh1,koh2}. In the present study these equations are implemented 
from an interaction Lagrangian density with four fermion contact interaction terms including
the isoscalar-scalar, isoscalar-vector, isovector-vector and isospace-space channels (for more
details see Refs.~\cite{nik08,nik14}),
\begin{equation}
\begin{split}
\mathcal{L} &= \bar{\psi}(i\gamma \cdot \partial - m)\psi  \\
 &-\frac{1}{2}\alpha_S(\rho) (\bar{\psi}\psi)(\bar{\psi}\psi) 
 -\frac{1}{2}\alpha_V(\rho) (\bar{\psi}\gamma^\mu \psi)(\bar{\psi}\gamma_\mu \psi) \\
 &-\frac{1}{2}\alpha_{TV}(\rho) (\bar{\psi}\vec{\tau}\gamma^\mu \psi)(\bar{\psi}\vec{\tau}\gamma_\mu \psi)
 \\
&-\frac{1}{2}\delta_S (\partial_\nu \bar{\psi} \psi)(\partial^\nu \bar{\psi} \psi) 
-e\bar{\psi}\gamma \cdot A \frac{1-\tau_3}{2}\psi .
\label{eq:PC-lag}
\end{split}
\end{equation}
In addition to the free-nucleon terms, the effective Lagrangian includes point coupling interaction
terms, coupling of protons to the electromagnetic field, and the derivative term accounting for
the leading effects of finite-range interactions necessary for a quantitative description of 
nuclear density distribution and radii. Starting from the microscopic density dependence of the
scalar and vector self-energies, the following functional form of the couplings is 
employed \cite{nik08,Entem.03},
\begin{equation}
\label{eq:alpha_i}
\alpha_i (\rho) = a_i + (b_i + c_i x)e^{-d_i x},\quad (i=S,V,TV),
\end{equation}
where $x=\rho/\rho_{0}$, and $\rho_{0}$ represents the nucleon density
in symmetric nuclear matter at saturation point. The point-coupling model includes 10 parameters ($a_{S}$, $b_{S}$, $c_{S}$, $d_{S}$, $a_{V}$, $b_{V}$, $d_{V}$
$b_{TV}$, $d_{TV}$ and $\delta_S$). For the description of open-shell nuclei, the relativistic
Hartree-Bogoliubov (RHB) model~\cite{nik14} is used and the pairing field is formulated using separable 
pairing force, which also includes two parameters for the pairing strength ($G_{p}$ and $G_{n}$) \cite{Tian}. 
More details and various implementations of the RHB model can be found in Refs. \cite{vre05,meng06}.
In the small amplitude limit, the collective excitations are described by the relativistic
(quasiparticle) random phase approximation (Q)RPA \cite{paar07}. In the present study, a unified computational framework of the RHB model~\cite{nik14} and self-consistent relativistic (Q)RPA~\cite{paar07} is established to constrain the 12 model parameters by minimizing the $\chi^{2}$ objective function \cite{rocc15}. 
In order to constrain the model parameters, the binding energies (34 nuclei), charge radii (26 nuclei) and mean pairing gaps (15 nuclei) of the selected open-shell nuclei are used along with the two observables on collective excitations: the constrained ISGMR energy ($E_{ISGMR}=\sqrt{m_{1}/m_{-1}}$) and dipole polarizability for $^{208}$Pb (see the supplementary material for details on the data set). Here, $m_{1}$ and $m_{-1}$ represent the energy-weighted moment and inverse energy-weighted moment of the strength distribution \cite{paar07}, respectively. The mean gap values for protons and neutrons are calculated using the five-point formula \cite{ben00}.  The adopted errors for the binding energies, charge radii and pairing gaps are taken as 1.0 MeV, 0.02 fm and 0.05 MeV, respectively. 
Recently, the dipole polarizability was measured in $^{208}$Pb using polarized proton inelastic scattering at extreme forward angles \cite{tami11}. After the subtraction of the quasi-deuteron effect from the experimental data, the dipole polarizability was obtained as 19.6$\pm$0.6 fm$^{3}$ in $^{208}$Pb \cite{roc15}. Several experimental studies were also performed to explore the ISGMR in $^{208}$Pb \cite{you04,you99,uc04,pat15,patel13}. Although the uncertainties for the measured ISGMR energies are small, there are differences in the excitation energies from different studies. Recently, the ISGMR energies were measured for $^{204,206,208}$Pb using inelastic $\alpha$-scattering at extremely forward angles, and the constrained ISGMR energy was found as 13.5$\pm$0.1 in $^{208}$Pb \cite{patel13}, whereas this energy was previously obtained as 14.18$\pm$0.11 MeV in the Texas A$\&$M experiment ~\cite{you99}. In optimizing the EDF, the constrained ISGMR energy is taken as 13.5 MeV \cite{patel13} and due to experimental uncertainties we adopted slightly large error (1.0\%) in the fitting protocol. The dipole polarizability is also taken as 19.6 fm$^{3}$ \cite{roc15} and the adopted error used is 0.5\%. As mentioned above, the implementation of the collective nuclear excitations in the fitting protocol is crucial in constraining both the isoscalar and isovector channels of the EDF's \cite{gar18,roc18}.

Using the observables introduced above, the ${\chi}^2$ minimization for the relativistic point coupling interaction is performed. This procedure is supplemented with the co-variance analysis that allows to determine statistical uncertainties of the model parameters
and other quantities, as well as relevant correlations between various properties~\cite{rein10,rocc15}.
Accordingly, the curvature matrix is determined at the ${\chi}^2$ minimum, $\mathcal{M}\equiv\partial_{p_i}\partial_{p_j}\chi^2$, where $p_i$ and $p_j$ denote the interaction parameters ($i,j=1,..,12$). Then, it is used to
estimate the uncertainties of the model parameters, using $\sigma(p_i) \equiv \sqrt{\left(\mathcal{M}^{-1}\right)_{ii}}$ relation \cite{rocc15}. The covariance between the two observables ($A,B$) is defined as ~\cite{rocc15}
\begin{equation}
cov(A,B)=cov(B,A)=\sum_{i,j=1}^{N}\left(\frac{\partial A}{\partial p_{i}}\right)\mathcal{M}_{ij}^{-1}\left(\frac{\partial B}{\partial p_{j}}\right),
\end{equation}
where the derivatives of the observables and the inverse of the curvature matrix are calculated at the $\chi^2$ minimum. The statistical uncertainty of any quantity of interest $A$ is calculated using $\sigma(A)=\sqrt{cov(A,A)}.$

The resulting DD-PCX parameterization with the respective statistical uncertainties are given in Table \ref{p}. 
The statistical uncertainties of the parameters are found to be small, indicating that the parameters of the interaction are well constrained.  
 
\begin{table}[ht]
\centering
\renewcommand\arraystretch{1.5}
\caption{Parameters of the DD-PCX interaction with the corresponding statistical uncertainties. The value of the nucleon mass is 939.0 MeV and the saturation density is set to 0.152 fm$^{-3}$.}
\begin{tabular}[t]{lcc}
\hline
parameters&DD-PCX&$\sigma$\\
\hline
$a_{s}$ (fm$^{2}$)      & -10.979243836  & 0.010808546   \\
$b_{s}$ (fm$^{2}$)      &  -9.038250910  & 0.023987420   \\
$c_{s}$ (fm$^{2}$)      &  -5.313008820  & 0.047152813   \\
$d_{s}$                 &   1.379087070  & 0.003900800   \\
$a_{v}$ (fm$^{2}$)      &   6.430144908  & 0.024888709   \\
$b_{v}$ (fm$^{2}$)      &   8.870626019  & 0.019549460   \\
$d_{v}$                 &   0.655310525  & 0.003073028   \\
$b_{tv}$ (fm$^{2}$)     &   2.963206854  & 0.092150525   \\
$d_{tv}$                &   1.309801417  & 0.053360277   \\
$\delta_{s}$ (fm$^{4}$) &  -0.878850922  & 0.004512226   \\
\hline
$G_{n}$ (MeV.fm$^{3}$)& -800.663126037 & 6.279054350   \\
$G_{p}$ (MeV.fm$^{3}$)& -773.776776597 & 4.003044910  \\
\hline
\label{p}
\end{tabular}
\end{table}%
Table \ref{nmp} shows the nuclear matter properties at the saturation for density-dependent point coupling interactions, DD-PCX (with uncertainties) and DD-PC1 \cite{nik08}, density dependent meson-exchange interaction DD-ME2 \cite{lal05}, and non-linear point coupling interaction PC-PK1 \cite{zha10}. These include the energy per nucleon $E/A$, the Dirac effective nucleon mass $m_D^{*}$~\cite{nik08}, the nuclear matter compression modulus $K_0$, the symmetry energy at saturation density $J$ and the slope of the symmetry energy at saturation $L$~\cite{roc18}. Compared to the DD-PCX, the DD-PC1, DD-ME2, and PC-PK1 interactions were established using different protocols. The DD-ME2 parameterization is based on the density-dependent meson-exchange interaction, constrained using the ground state properties of spherical nuclei \cite{lal05}, whereas the DD-PC1 interaction is based on the point-coupling model, and optimized using the binding energies of deformed nuclei \cite{nik08}. In addition, selected nuclear matter properties were fixed in both interactions, and no data on excitations have been used in the ${\chi}^2$ minimization. In Ref.~\cite{zha10} the non-linear point coupling interaction PC-PK1 has been 
constrained by fitting to observables of 60 selected spherical nuclei, including the binding energies, charge radii, and empirical pairing gaps, and no constraints on the nuclear matter properties have been used in the fitting protocol.
Due to the implementation of the nuclear excitations for $^{208}$Pb in constraining the DD-PCX interaction,
we find that the incompressibility ($K_{0}$) and symmetry energy parameters ($J$ and $L$) at saturation density
are lower than for the DD-PC1, DD-ME2, and PC-PK1 effective interactions.
The respective uncertainties are also found to be small, indicating that within the fitting protocol employed both the isoscalar and isovector channels of the interaction are tightly constrained. The DD-PCX interaction values for $J$ and $L$ are in agreement with the suggested values from previous studies~\cite{roc18}. The compression modulus $K_{0}$ is also found at around 213 MeV, which is lower in comparison to other relativistic interactions~\cite{nik08,lal05}. We realize that small $K_0$ is a direct
consequence of the more recent experimental data ($E_{ISGMR}$=13.5$\pm$0.1 MeV)~\cite{patel13} used in the fitting protocol, and the implementation of the data from Texas A$\&$M experiment ($E_{ISGMR}$=14.18$\pm$0.11 MeV) \cite{you99} would lead toward the higher value of $K_{0}$. Clearly, resolving the ambiguities between different experimental studies on the ISGMR data for $^{208}$Pb is essential for constraining the nuclear matter incompressibility.

\begin{table}[ht]
\centering
\renewcommand\arraystretch{1.5}
\caption{The saturation properties of nuclear matter for the DD-PCX, DD-PC1
\cite{nik08}, DD-ME2 \cite{lal05}, and PC-PK1 \cite{zha10} interactions.}
\begin{tabular}[t]{lcccc}
\hline
&DD-PCX&DD-PC1&DD-ME2&PC-PK1\\
\hline
E/A (MeV)  & -16.026 $\pm$ 0.018 & -16.06 & -16.14 & -16.12       \\
$m^{*}_{D}/m$  & 0.5598 $\pm$ 0.0008 & 0.580 & 0.572 & 0.590   \\
$K_{0}$ (MeV)   & 213.03 $\pm$ 3.54  & 230.0 & 250.89 & 238.0    \\
$J$ (MeV)   & 31.12 $\pm$ 0.32& 33.0 & 32.30& 35.60           \\
$L$ (MeV)  & 46.32 $\pm$ 1.68 & 70.0 &  51.26 & 113.0          \\
\hline
\label{nmp}
\end{tabular}
\end{table}%

\begin{figure}[!ht]
 \begin{center}
\includegraphics[width=\linewidth,clip=true]{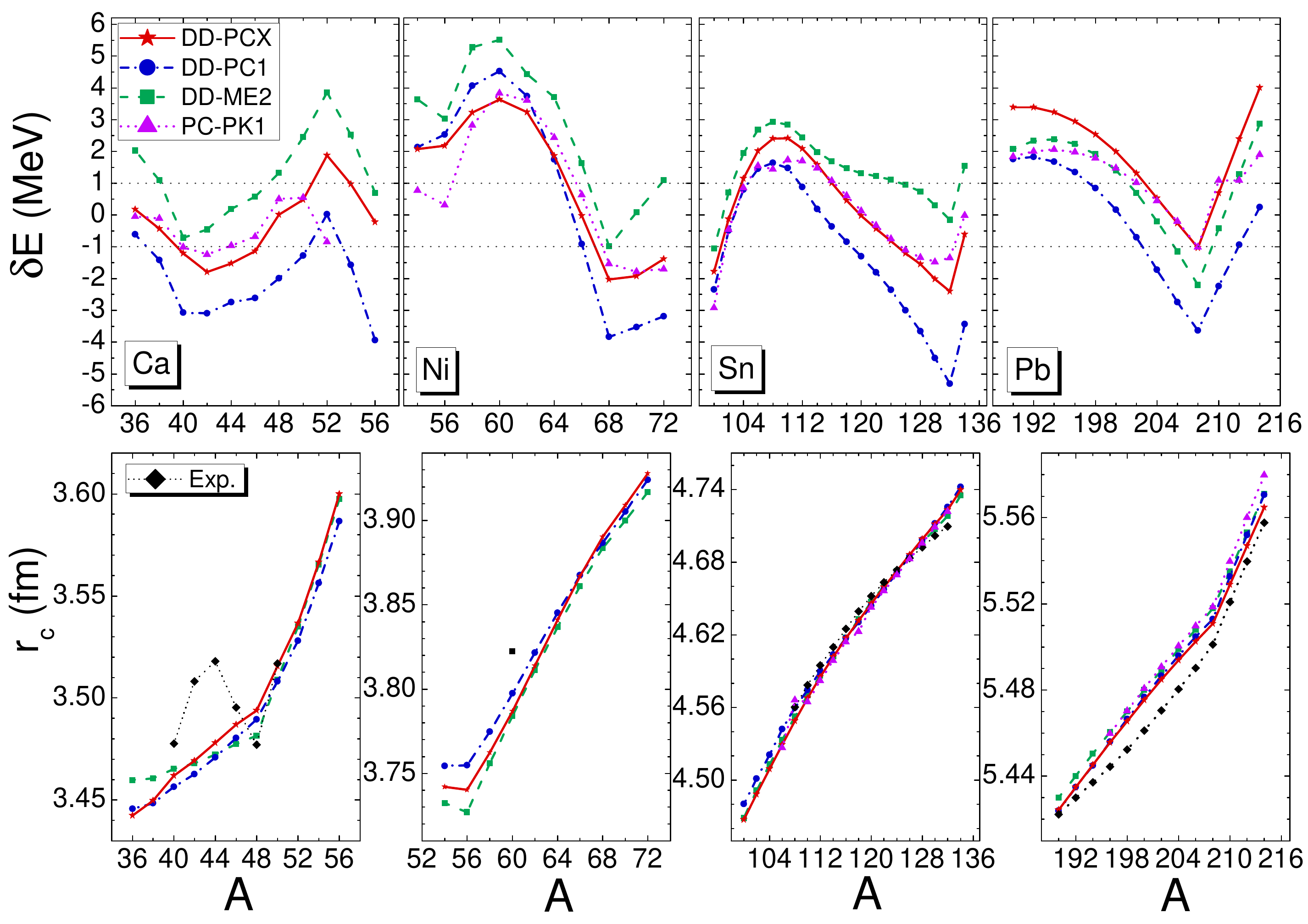}
 \end{center}
 \caption{Upper panel: the difference between the experimental \cite{mass} and calculated binding energies for Ca, Ni, Sn and Pb isotopes. Lower panel: the calculated charge radii in comparison with the experimental values \cite{charge} . The calculations are performed using the DD-PCX, DD-PC1 \cite{nik08}, and DD-ME2 \cite{lal05} interactions. Available data for the PC-PK1 interaction from Refs. \cite{zha10,zha10b} are also shown.} 
  \label{be}
\end{figure}

In the following, the performance of the DD-PCX interaction in description of various nuclear ground state and excitation properties is presented.
Figure \ref{be} shows the differences between the experimental \cite{mass} and calculated binding energies (upper panel) and the charge radii for Ca, Ni, Sn and Pb isotopes (lower panel) for the DD-PCX interaction. For comparison, the results are also shown for the density dependent point coupling interaction DD-PC1 \cite{nik08}, density dependent meson-exchange interaction DD-ME2 \cite{lal05}, and non-linear point coupling interaction PC-PK1~\cite{zha10,zha10b}. In this work, the calculations are performed using the RHB model with separable pairing force and spherical symmetry is assumed. 
It is seen that all interactions under consideration provide a reasonable description of the binding
energies and produce similar isotopic dependencies. Compared to the DD-PC1 and DD-ME2, the newly parametrized DD-PCX interaction seems to be more successful in the predictions of the experimental binding energies of spherical nuclei. Since the pairing parameters ($G_{n}$, $G_{p}$) are also included in the fitting protocol, the success of the DD-PCX can also be related to the better reproduction of pairing properties of nuclei as discussed below. In comparison to the experimental data~\cite{charge},
the charge radii are also accurately reproduced, with a few exceptions for Ca isotopes and $^{60}$Ni. 
In order to assess a more general overview about the performance of the DD-PCX interaction, we calculate the root mean square error ($\Delta$) and root mean square relative error ($\delta$) in percentage for binding energies, charge radii, and mean pairing gaps for a set of nuclei (see supplementary material for the list of nuclei), and the results are given in Table \ref{mass}. It is seen that the resulting binding energies and mean gap values are better reproduced using the DD-PCX interaction, while the deviations for the charge radii are similar in all cases. An extended list with the calculated binding energies and charge radii for the selected nuclei using the same interactions is also given in the supplemental material with the corresponding experimental data.

 \begin{table}[!htbp] \centering
 \caption{The root mean square error ($\Delta$) and root mean square relative error ($\delta$) in percentage for the binding energies (B.E.) (MeV), charge radii ($r_{c}$) (fm) and mean gaps (MeV) for a set of spherical nuclei, using the relativistic DD-PCX, DD-PC1 and DD-ME2 interactions. The numbers of the nuclei considered in the calculations are given in parentheses.}
 \label{}
 \begin{tabular}{@{\extracolsep{5pt}}lccccc} 
 \\[-1.8ex]\hline 
 \hline \\[-1.8ex] 
    &  & B.E. & $r_{c}$ & Mean Gap \\
Interaction && (65) & (46) & (56)\\
 \hline \\[-1.8ex]  
 \multirow{ 2 }{*}{ DD-PCX }  &  $\Delta$    &  1.38 MeV       &  0.016 fm     & 0.18 MeV         \\
 \hhline{~~~~~}               &  $\delta$    &  0.21\%     &  0.47\%    & 15.40\%          \\
 \hline \\[-1.8ex]  
\multirow{ 2 }{*}{ DD-PC1 }   &  $\Delta$    &  3.05 MeV      &  0.017 fm    & 0.29  MeV         \\
 \hhline{~~~~~}               &  $\delta$    &  0.48\%    &  0.49\%    & 21.73\%          \\
 \hline \\[-1.8ex] 
\multirow{ 2 }{*}{ DD-ME2 }   &  $\Delta$    &  2.08 MeV      &  0.016 fm    & 0.35 MeV        \\
 \hhline{~~~~~}               &  $\delta$    &  0.27\%    &  0.44\%   & 26.13\%            \\
 \hline \\[-1.8ex] 
 \end{tabular}
\label{mass}
 \end{table}

The predictive power of the DD-PCX interaction is also tested on the nuclear excitations of nuclei. 
Table \ref{gmr} shows the constrained ISGMR energies for $^{90}$Zr, $^{120}$Sn, $^{208}$Pb calculated using various relativistic interactions, and compared to the experimental results ~\cite{gup16,li10, kr15,patel13,you99}. In this part, we should mention that the DD-ME2 calculations are performed using the finite range Gogny interaction D1S in the particle-particle channel for open-shell nuclei~\cite{paar03}, while separable pairing is used with the DD-PCX and DD-PC1 interactions. Among the relativistic interactions, the DD-ME2 (DD-PC1) interaction predicts the lowest (highest) values for the constrained ISGMR energies. Considering the point coupling interactions, the table clearly 
demonstrates the relevance
of including the constraint on the ISGMR excitation energy for $^{208}$Pb in order to provide a reasonable 
description of the ISGMR energies in all nuclei under consideration. Using the DD-PCX interaction, the calculated values for $^{90}$Zr and $^{120}$Sn are slightly above the
experimental values and further fine-tuning
of the interaction may be achieved by considering the ISGMR energies of additional nuclei and using smaller adopted error within the $\chi^2$ minimization of the interaction. However, the softness of the Sn nuclei still represents an open question, and their inclusion in constraining the interaction may cause difficulties.

\begin{table}[ht]
\centering
\renewcommand\arraystretch{1.5}
\caption{The constrained ISGMR energies (in MeV) for $^{90}$Zr, $^{120}$Sn and $^{208}$Pb in comparison with the experimental data.}
\begin{tabular}[t]{lcccc}
\hline
&Exp.&DD-PCX&DD-PC1&DD-ME2\\
\hline
$^{90}$Zr   &17.58$_{-0.04}^{+0.06}$\cite{kr15} & 18.00 $\pm$0.10 & 18.83  &  17.80     \\
            &17.66$\pm$0.07 \cite{gup16}        &       &        &            \\
$^{120}$Sn  &15.5$\pm$0.1 \cite{li10}           & 16.18 $\pm$0.09 & 16.92  &  16.07     \\
$^{208}$Pb  &13.5$\pm$0.1 \cite{patel13}        & 13.66 $\pm$0.08 & 14.22  &  13.49      \\
            &14.18$\pm$0.11 \cite{you99}        &       &        &             \\
\hline
\label{gmr}
\end{tabular}
\end{table}%
\begin{figure}[!ht]
 \begin{center}
\includegraphics[width=\linewidth,clip=true]{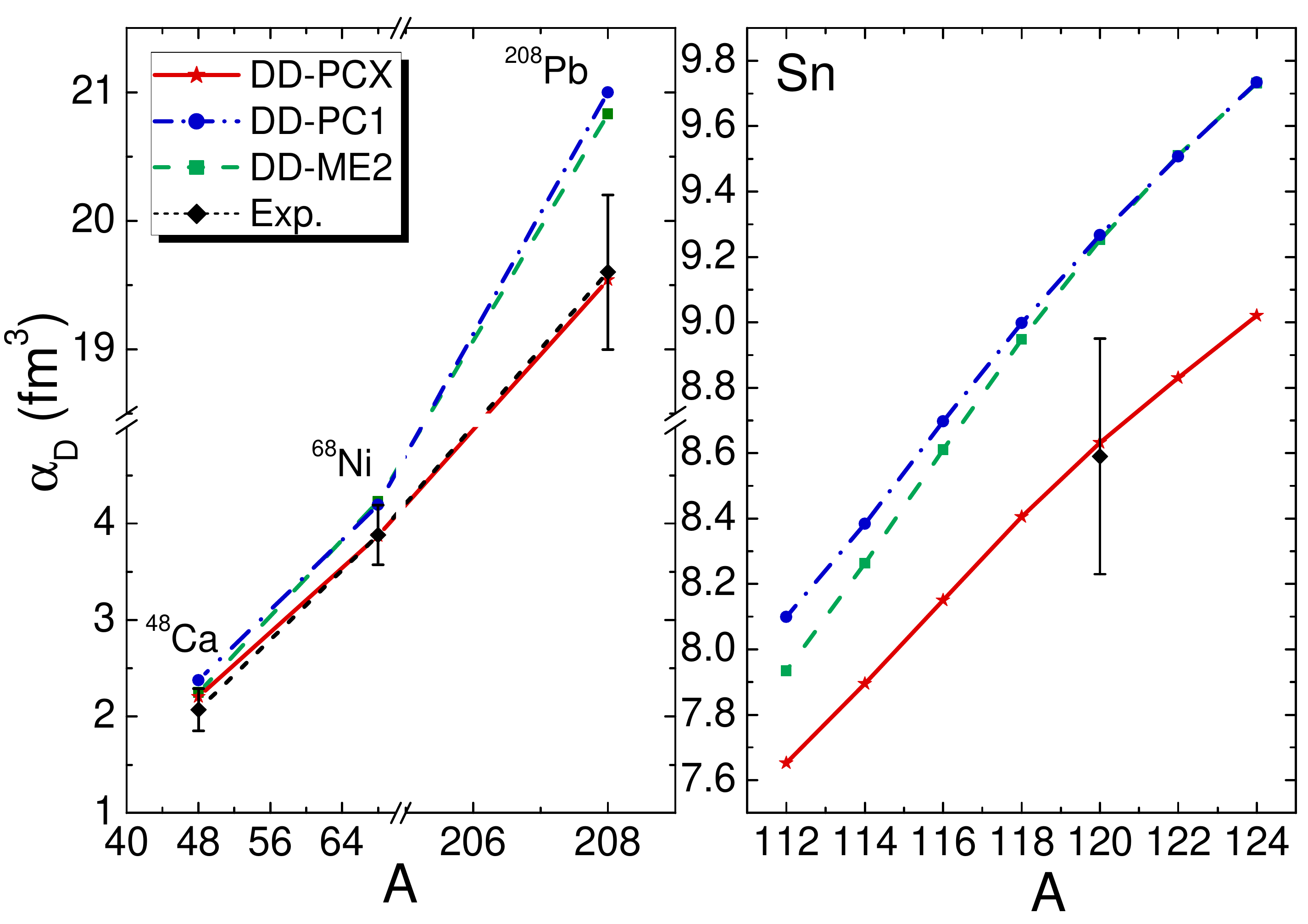}
 \end{center}
 \caption{The dipole polarizabilities for $^{48}$Ca, $^{68}$Ni and $^{208}$Pb (left panel) and tin isotopic chain (right panel). The calculations are performed using RHB+(Q)RPA with DD-PCX, DD-PC1 and DD-ME2 interactions. The experimental data are taken from Refs. \cite{bir17,ros13,has15,tami11, roc15}.} 
  \label{ad}
\end{figure}

Due to the empirical constraint imposed on the dipole polarizability for $^{208}$Pb, it is expected that the isovector 
channel of the DD-PCX interaction is improved in comparison to other effective interactions, 
as illustrated in a few following examples. Figure \ref{ad} shows the dipole 
polarizabilities for $^{48}$Ca, $^{68}$Ni and $^{208}$Pb (left panel) and Sn isotopic chain (right panel). The calculations are performed using the DD-PCX, DD-PC1 and DD-ME2 interactions, and the available experimental data are also provided~\cite{bir17,ros13,has15,tami11,roc15}. As shown in Fig. \ref{ad}, only the DD-PCX interaction systematically reproduces the experimental results on $\alpha_D$ for all nuclei under consideration, while the DD-ME2 and DD-PC1 interactions mainly overestimate the measured values. Although in constraining the functional we only use the dipole polarizability for $^{208}$Pb, the DD-PCX interaction is also successful in the prediction of the dipole polarizabilities for other nuclei, as expected due to the strong correlation between the dipole polarizabilities in different nuclei~\cite{roc15}.

\begin{figure}[!ht]
\begin{center}
\includegraphics[width=\linewidth,clip=true]{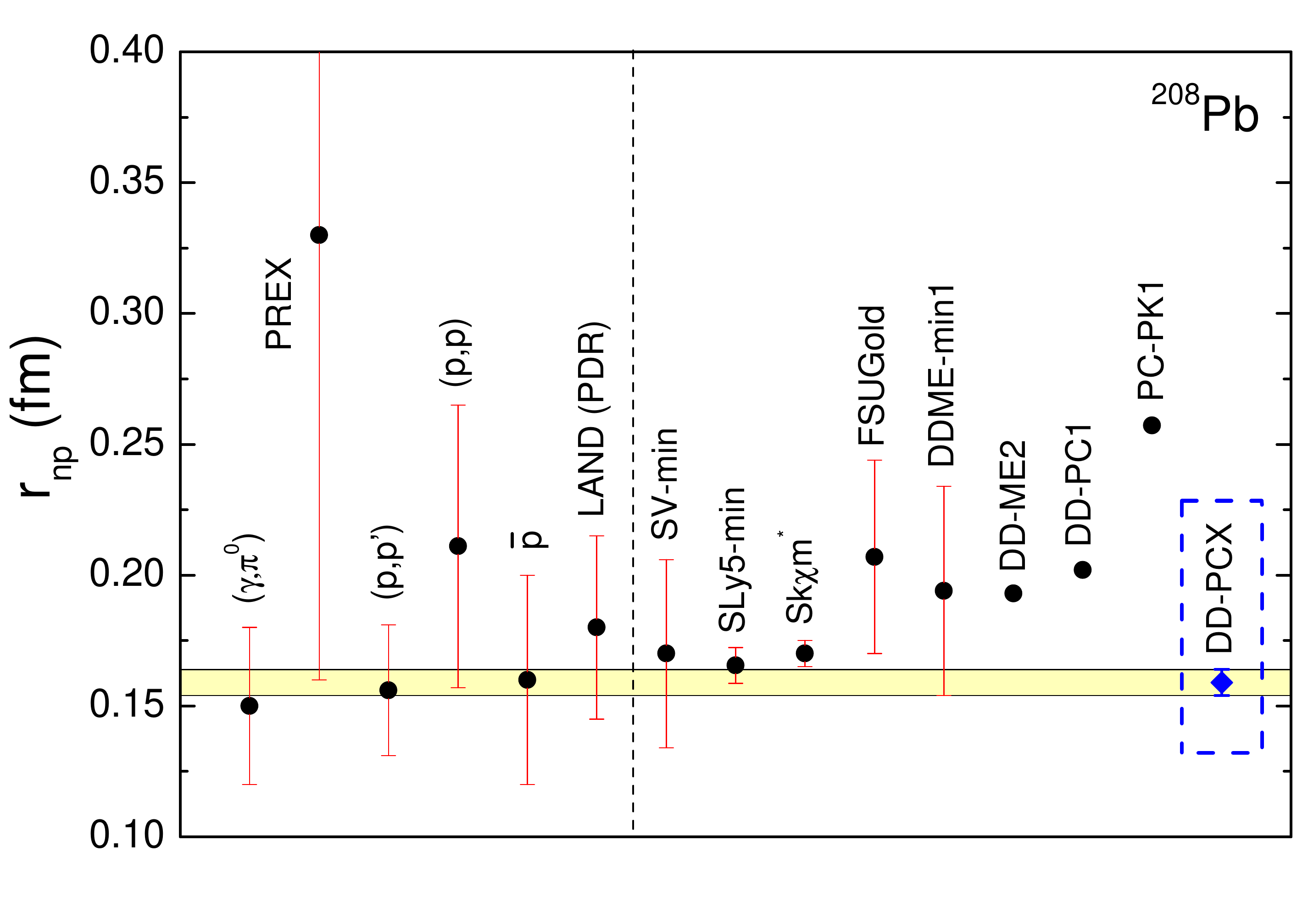}
\end{center}
 \caption{The neutron skin thickness for $^{208}$Pb predicted by different experiments and nuclear energy density functionals. The experimental data are taken from ($\gamma$,$\pi^{0}$) \cite{tar14}, PREX \cite{prex12}, ($p,p'$) \cite{tami11}, ($p,p$) \cite{pp11}, $\overline{p}$ \cite{antip07}, LAND(PDR) \cite{klim07}. The calculations with the non-relativistic interactions: SV-min \cite{klup09}, SLy5-min~\cite{rocc15}, Sk$\chi$\textit{m}* \cite{zhen18}, and relativistic interactions: FSUGold \cite{todd05}, DDME-min \cite{rocc15}, DD-ME2 \cite{lal05}, DD-PC1 \cite{nik08}, PC-PK1 \cite{zha10}.} 
  \label{r_np}
\end{figure}

Another important isovector property to be considered is the neutron skin thickness $r_{np}$.
In Fig.~\ref{r_np}, the neutron skin thickness for $^{208}$Pb is shown for a set of relativistic~\cite{rocc15,zha10,todd05,lal05,nik08} and non-relativistic calculations~\cite{klup09,rocc15,zhen18} along with the experimental data from Refs.~\cite{tar14,prex12,tami11,pp11,antip07,klim07}. Using the DD-PCX interaction, the neutron skin thickness  for $^{208}$Pb is predicted as $r_{np}$=0.159$\pm$0.005 fm, and the (yellow) band denotes the calculated statistical uncertainty.
Considering the experimental results, it is clear that there are discrepancies in the predictions for $r_{np}$ values in $^{208}$Pb. 
The non-relativistic functionals predict lower values for the neutron skin thickness compared to the relativistic ones, with 
the exception of the new effective interaction DD-PCX. Among the relativistic interactions, the DD-PCX prediction provides the lowest neutron skin thickness for $^{208}$Pb, which is also in a good agreement with the majority of the experimental data. The neutron-skin thicknesses of neutron-rich $^{48}$Ca and $^{132}$Sn nuclei are also calculated as 0.172$\pm$0.003 fm and 0.218$\pm$0.005 fm, respectively. We find that the calculated neutron skin thicknesses are also in a good agreement with the ``model-averaged" results in Ref. \cite{pie12}. The results demonstrate that the DD-PCX interaction can also be used to make reliable predictions for the neutron skin thicknesses of other nuclei.

%

In conclusion, we have established a unified theoretical framework to constrain the relativistic nuclear EDF, based on the RHB plus (Q)RPA, supplemented with the co-variance analysis, using not only the nuclear ground state properties, but also relevant properties on collective nuclear excitations. The relativistic point-coupling interaction DD-PCX introduced in this work, represents the first effective interaction that is constrained using the binding energies, charge radii and pairing gaps, together with a direct implementation of the ISGMR energy and dipole polarizability in the $\chi^2$ 
minimization. In comparison to the previous studies, where the functionals have the properties
of the nuclear symmetry energy either unconstrained, constrained by the pseudo-observables on nuclear matter
that are often arbitrary, or validated by the data after the parameters have been determined, 
the present study implements directly genuine observables on collective excitations to optimize the effective nuclear interaction. The success of the DD-PCX interaction in the predictions of the dipole polarizabilities and neutron skin thicknesses in other nuclei not used in optimizing the model parameters validates the isovector channel of the functional and the respective symmetry energy properties. The present analysis clearly demonstrates the relevance of accurate measurements of the nuclear collective phenomena, as well as the necessity to resolve the ambiguities in the existing data from different experiments, both in the isoscalar (e.g., the ISGMR energy in $^{208}$Pb~\cite{you99,patel13} and isovector sectors for constraining modern nuclear energy density functionals.

We thank T. Nik\v{s}i\'{c} for stimulating discussions while preparing this paper.
This work is supported by the Croatian Science Foundation under the project Structure and Dynamics
of Exotic Femtosystems (IP-2014-09-9159) and by the QuantiXLie Centre of Excellence, a project co financed by the Croatian
Government and European Union through the European Regional Development Fund, the Competitiveness and Cohesion
Operational Programme (KK.01.1.1.01). E. Y. also acknowledges financial support from the Scientific
and Technological Research Council of Turkey (T\"{U}B\.{I}TAK) BIDEB-2219 Postdoctoral Research program.

\end{document}